\documentclass[twocolumn]{revtex4}
\usepackage{dcolumn}
\usepackage{bm}
\usepackage{graphicx}
\usepackage{amsmath}
\usepackage{latexsym}
\usepackage{amsfonts}
\usepackage{amssymb}
\usepackage{array}
\usepackage{epsfig}
\usepackage{times}
\usepackage{txfonts}
\usepackage{epstopdf}
\usepackage{pifont}
\usepackage{dsfont}
\usepackage{amscd}
\usepackage{amsfonts}
\usepackage{textcomp}

\newcommand{\be}{\begin{equation}}
\newcommand{\ee}{\end{equation}}
\newcommand{\bea}{\begin{eqnarray}}
\newcommand{\eea}{\end{eqnarray}}

\begin{document}
\title{High-speed photon correlation monitoring of amplified quantum noise by chaos using deep-learning balanced homodyne detection}

\author{Yanqiang Guo$^{1,2}$, Zinan Hu$^{1,2}$, Jianchao Zhang$^{1}$, Chenyu Zhu$^{1}$, and Xiaomin Guo$^{1,}$}
\altaffiliation{Electronic mail: guoxiaomin@tyut.edu.cn}
\affiliation{$^1$Key Laboratory of Advanced Transducers and Intelligent Control System, Ministry of Education, College of Physics, Taiyuan University of Technology, Taiyuan 030024, China\\
$^2$State Key Laboratory of Cryptology, Beijing 100878, China}%
\date{\today}

\begin{abstract}
Precision experimental determination of photon correlation requires the massive amounts of data and extensive measurement time. We present a technique to monitor second-order photon correlation $g^{(2)}(0)$ of amplified quantum noise based on wideband balanced homodyne detection and deep-learning acceleration. The quantum noise is effectively amplified by an injection of weak chaotic laser and the $g^{(2)}(0)$ of the amplified quantum noise is measured with a real-time sample rate of 1.4 GHz. We also exploit a photon correlation convolutional neural network accelerating correlation data using a few quadrature fluctuations to perform a parallel processing of the $g^{(2)}(0)$ for various chaos injection intensities and effective bandwidths. The deep-learning method accelerates the $g^{(2)}(0)$ experimental acquisition with a high accuracy, estimating 6107 sets of photon correlation data with a mean square error of 0.002 in 22 seconds and achieving a three orders of magnitude acceleration in data acquisition time. This technique contributes to a high-speed and precision coherence evaluation of entropy source in secure communication and quantum imaging.
\end{abstract}

\maketitle

Quantum noise arising from the minimum uncertainty nature of quantum mechanics is of great importance for quantum precision measurement and quantum cryptography\cite{Clerk10,Xiao87,Gisin02,Chen21,Liu23}. As an unpredictable, irreproducible, and secure state, quantum noise has been successfully applied to high-speed secure communication and quantum random number generation \cite{Braunstein05,Gabriel10,Symul11,Guo19}, but quantum noise with low fluctuation intensity is not easy to extract efficiently, and high-speed cryptographic applications require entropy amplification of quantum noise to ensure high security and fast key generation rate \cite{Guo182,Guo21}. Meanwhile, characterizing the statistical properties of noise source is fundamental for the improvement of key generation rate and information security \cite{Guo181}. Second-order photon correlation $g^{(2)}$ is a hallmark characterization to determine the quantum statistics and photon emission process of light fields \cite{Glauber63}. The pioneering experiment of the photon correlation measurement is conducted by Hanbury Brown and Twiss (HBT) \cite{Brown56}, and afterwards the $g^{(2)}$ measurement with advanced single photon detection is harnessed in many applications, such as quantum imaging \cite{Ryczkowski16,Yao18,Li13}, quantum computing \cite{o07,zhong20}, spectral measurement \cite{Goda08,Hlousek19} and quantum communication \cite{Kuhn02,Deng16,Li23,Ma22,Ou22}. Moreover, the photon correlation $g^{(2)}(0)$ at zero time delay has been a standard metric for categorizing light fields \cite{Guo22,Guo12}. However, the photon correlation measurement using single photon detection requires recording massive amounts of data and costing extensive analysis resources, due to weak signal light intensities, low detection bandwidth and other nonideal experimental conditions \cite{Luders18,Guo20}. The measurement time generally ranges from minutes to hours \cite{Schneeloch19} and even days. High-speed photon correlation measurement of light sources remains to be explored and is important for security monitoring in secure communication and real-time quantum random number generation.
\\ \indent The enhancement of the $g^{(2)}$ measurement bandwidth would place a higher demand on the processing capacity. Deep learning can be a candidate to accurately extract features and efficiently achieve data analysis, and the technique has been used to accelerate optical quantum experiment \cite{Hentschel10}. In recent years, deep learning has made significant achievements in the fields of identifying optical fields \cite{You20}, classifying quantum states \cite{Navarathna21}, predicting phonon blockade \cite{Zeng21}, and verifying quantum dynamics \cite{Xu21}. It is possible to reconstruct one set of measured photon events of thermal noise in traditional HBT interferometry with single photon detectors \cite{Cortes20}. However, the limited bandwidth of single photon detection makes the $g^{(2)}(0)$ dynamic measurement of noise sources difficult, and simultaneous speed-up estimation multiple sets of the photon correlation data remains a key challenge.
\\ \indent In this letter, we experimentally investigate the second-order photon correlation $g^{(2)}(0)$ of amplified quantum noise by chaos based on wideband balanced homodyne detection (BHD). A deep-learning convolutional neural network is developed to process the massive $g^{2}(0)$ data in parallel. We accelerate the $g^{(2)}(0)$ experimental acquisition of amplified quantum noise with a negligible mean square error, and high-resolution maps of the $g^{(2)}(0)$ are measured over a wide range of chaos injection intensity-bandwidth parameter space. We also compare our speed-up method with other deep-learning estimation. The results provide a solid evidence for accelerating photon correlation measurement involving dynamic amplified quantum noise.
\\ \indent Unlike traditional scheme of single photon detection, we develop a high-speed measurement for photon correlation of amplified quantum noise by chaos using wideband balanced homodyne detection. In the continuous-variable BHD strategy, a strong local oscillator (LO) and a weak chaos signal of orthogonal linear polarization are combined in the polarizing beam splitter (PBS). The group of half-wave plate (HWP) and PBS is used to accurately balance the two light intensities. Then the quadrature fluctuations of the input state can be acquired, and the operator of detecting a photon on the $+(-)$ photodiode is described as \cite{McAlister97}
\begin{equation}
\hat{a}_{ \pm}=\frac{1}{\sqrt{2}}\left(e^{i \varphi} \hat{a}_{L O} \pm \hat{a}_{S}\right),
    \label{eq1}
 \end{equation}
 where $\hat{a}_{LO}$ and  $\hat{a}_{S}$ are the photon annihilation operators for the LO and the signal, and $\varphi$ is the relative phase between the two lights. When the LO is a strong coherent light and the signal is a weak light, the difference photocurrent $\Delta I$ can be expressed as:
 \begin{equation}
    \Delta I =\hat{a}_{+}^{\dagger} \hat{a}_{+}-\hat{a}_{-}^{\dagger} \hat{a}_{-}=%
    \sqrt{N_{L O}}\left(e^{i\phi} \hat{a}_{s}^{\dagger}+e^{-i \phi} \hat{a}%
    _{s}\right).
    \label{eq2}
\end{equation}
Here $N_{L O}$ is the photon number of LO and ${N_{L O}}=\left\langle \Delta I^{2}\right\rangle-\left\langle \Delta I\right\rangle^{2}$. The moments $\left\langle \Delta I^{2}\right\rangle$ and $\left\langle \Delta I^{4}\right\rangle$ can be obtained as follows:
\begin{eqnarray}
    \left\langle \Delta I^{2}\right\rangle&=&{\sqrt{N_{L O}}}^{2}\left(2\left\langle\hat{a}_{s}^{\dagger} \hat{a}_{s}\right\rangle+1\right)\nonumber\\
    & &\Rightarrow\left\langle\hat{a}_{s}^{\dagger} \hat{a}_{s}\right\rangle
    =\frac{1}{2}\left(\frac{\left\langle \Delta I^{2}\right\rangle}{N_{L O}}-1\right),
    \label{eq3}
\end{eqnarray}

\begin{eqnarray}
 \left\langle \Delta I^{4}\right\rangle&=&{\sqrt{N_{L O}}}^{4}\left(6\left\langle\hat{a}_{s}^{\dagger} \hat{a}_{s}^{\dagger} \hat{a}_{s} \hat{a}_{s}\right\rangle+12\left\langle\hat{a}_{s}^{\dagger} \hat{a}_{s}\right\rangle+3\right)\nonumber\\
& &\Rightarrow\left\langle\hat{a}_{s}^{\dagger} \hat{a}_{s}^{\dagger} \hat{a}_{s} \hat{a}_{s}\right\rangle=\frac{\left\langle \Delta I^{4}\right\rangle}{6 N_{L O}^{2}}-\frac{\left\langle \Delta I^{2}\right\rangle}{N_{L O}}+\frac{1}{2}.
    \label{eq4}
\end{eqnarray}

The measured second-order photon correlation $g^{(2)}$(0) is derived by the moments $\left\langle \Delta I^{2}\right\rangle$ and $\left\langle \Delta I^{4}\right\rangle$,
\begin{eqnarray}
    g^{(2)}(0)&=&\frac{\left\langle\hat{a}_{s}^{\dagger} \hat{a}_{s}^{\dagger} \hat{a}_{s} \hat{a}_{s}\right\rangle}{\left\langle\hat{a}_{s}^{\dagger} \hat{a}_{s}\right\rangle^{2}}=\frac{\frac{2\left\langle \Delta I^{4}\right\rangle}{3 N_{L O}^{2}}-4 \frac{\left\langle \Delta I^{2}\right\rangle}{N_{L O}}+2}{\left(\frac{\left\langle \Delta I^{2}\right\rangle}{N_{L O}}-1\right)^{2}}\nonumber\\
    &=&\frac{\frac{2}{3}\left\langle \Delta I^{4}\right\rangle-4 N_{L O}\left\langle \Delta I^{2}\right\rangle+2 N^{2}{ }_{L O}{ }}{\left(\left\langle \Delta I^{2}\right\rangle-N_{L O}\right)^{2}}.
    \label{eq5}
\end{eqnarray}

\begin{figure}[htb]
\includegraphics[width=0.48\textwidth]{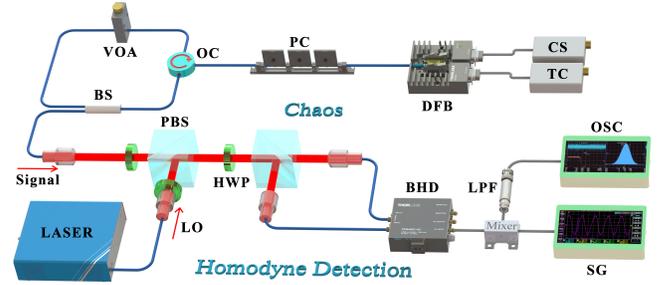}
\caption{\label{fig:1}Schematic illustration of the experimental setup. CS: current source; TC: temperature controller ; DFB: semiconductor laser; PC: polarization controller; OC: optical circulator; VOA: attenuator; BS: 20/80 beam splitter; Laser: local oscillator laser; PBS: polarizing beam splitter; HWP: half-wave plate; BHD: balanced homodyne detection; LPF: low pass filter; OSC: oscilloscope; SG: signal generator.}
\end{figure}

Based on the above theoretical analysis, we experimentally investigate the second-order photon correlation $g^{(2)}(0)$ of amplified quantum noise by chaos based on the high-speed BHD. The high entropy source and its rapid characterization are closely coupled into a whole scheme. The experiment schematic is shown in Fig.~\ref{fig:1}. The experimental setup consists of two main parts: one part is the preparation of amplified quantum noise (AQN) by a chaotic laser and the other is the high-speed $g^{(2)}(0)$ measurement of AQN based on a wideband BHD. In the first part, the distributed feedback (DFB) semiconductor laser is stabilized at 1550 nm using a temperature control (TC) and a current source (CS) with the accuracies of 0.01 ${{}^\circ}C$ and 0.1 mA, respectively. The threshold current of the DFB laser $I_{th}$ is 6 mA. The laser connects to a polarization controller (PC) and its output passes through a optical circulator (OC) and a 20:80 beam splitter (BS). The principal output of the light is fed back to the DFB laser and the variable optical attenuator (VOA) is used to accurately control the feedback intensities \cite{Guo21}. The remaining 20$\%$ of the light enters into the balanced homodyne detection system for the photon correlation measurement. In the other part, a laser source with a wavelength of 1550 nm outputs a single-mode continuous light, which acts as the LO. The interference of the LO and chaos occurs at the first PBS and achieves high visibility. Then the interference light arrives at the second group of HWP and PBS and is divided into two equal beams. The BHD (Thorlabs PDB480C, 1.6 GHz) is devoted to acquiring the quadrature fluctuations of the AQN. The quantum shot noise is amplified effectively by the injection of the chaotic laser. The AQN is mixed down with a radio frequency and filtered using a low pass filter (LPF) of 1.4 GHz cut-off frequency. The time series and power spectrum of the AQN are recorded by a real-time oscilloscope with 36 GHz bandwidth (OSC, Lecroy, LabMaster10-36Zi) and a 26.5 GHz spectrum analyzer (Agilent N9020A), respectively.

\begin{figure}[htb]
\includegraphics[width=0.48\textwidth]{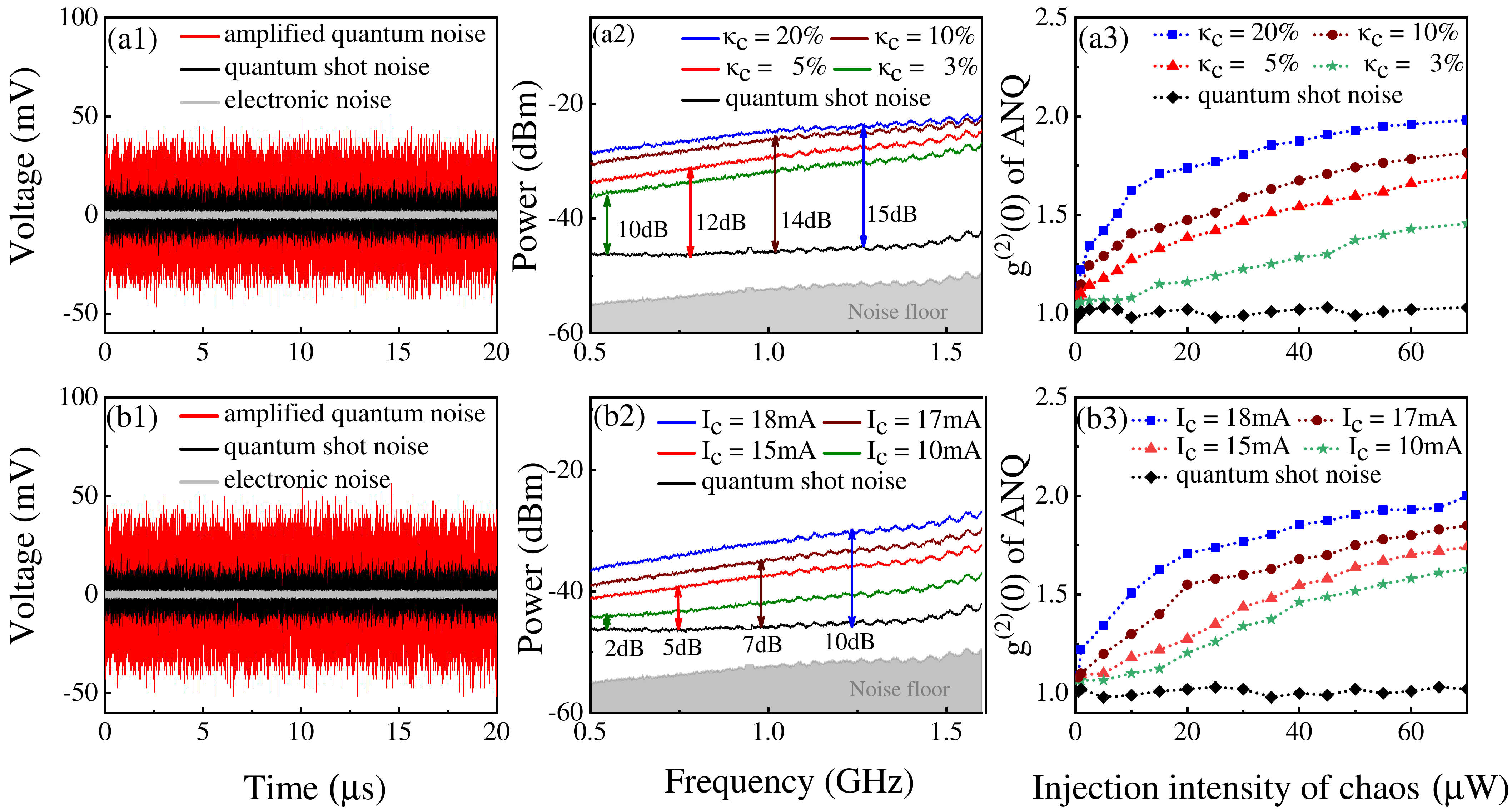}
\caption{\label{fig:2}Time series, power spectrum, and $g^{(2)}(0)$ of amplified quantum noise (a) for feedback strengths $\kappa_{c}=3\%$, $\kappa_{c}=5\%$, $\kappa_{c}=10\%$, $\kappa_{c}=20\%$ and (b) for $I_{c}=10mA$, $I_{c}=15mA$, $I_{c}=17mA$, and $I_{c}=18mA$.}
\end{figure}

The intensity of the AQN is stronger than that of quantum shot noise, as shown in Fig.~\ref{fig:2}(a1). The quantum shot noise is prepared through homodyne quadrature measurement \cite{Gabriel10,Guo19}, when the LO and the vacuum state interfere on a symmetric beamsplitter. The signal-to-noise ratio of the quantum shot noise is significantly improved by the chaotic amplification. Furthermore, the amplification effect is investigated for various feedback strengths and bias currents in the chaotic laser. The feedback strength is defined as the power ratio of the returned optical intensity to the total output intensity of the DFB laser. In Fig.~\ref{fig:2}(a2), the power spectrum of the AQN increases as the feedback strength $\kappa_{c}$ of the chaotic laser increases. For the bias current of the chaotic laser $I_{c}=18mA$, the AQN spectrum is higher than 10 dB, 12 dB, 14 dB, and 15 dB above the shot noise spectrum for $\kappa_{c}$ values of $3\%$, $5\%$, $10\%$, and $20\%$, respectively. Then we can obtain the second-order photon correlation $g^{(2)}(0)$ of the AQN. For various chaotic feedback strengths $\kappa_{c}$, the $g^{(2)}(0)$ of the AQN increases from 1 to 2 as the injection intensity of the chaotic laser increases, as shown in Fig.~\ref{fig:2}(a3). The results indicate that the photon bunching effect of the AQN is enhanced as the injection intensity of chaotic laser.

Moreover, the amplification effect associated with the bias current of the chaotic laser on the quantum noise is investigated experimentally. For $I_{c}=17mA$ and $\kappa_{c}=20\%$, the time series of AQN is shown in Fig.~\ref{fig:2}(b1). When the feedback strength is $\kappa_{c}=20\%$, the power spectrum of the AQN is more than 2 dB, 5 dB, 7 dB, and 10 dB above that of the shot noise for $I_{c}$ values of 10mA, 15mA, 17mA, and 18mA, as shown in Fig.~\ref{fig:2}(b2). Compared to no-injection case, the significant power enhancements of the quantum noise are achieved by varying the feedback strength and bias current of the injecting chaotic signal. The results of Fig.~\ref{fig:2}(b3) indicates that the $g^{(2)}(0)$ of the ANQ increases as the injection intensity and the bias current of the chaotic laser increase. The controllable amplification and coherence of the quantum noise is achieved by the chaotic laser injection.

\begin{figure}[htb]
\includegraphics[width=0.48\textwidth]{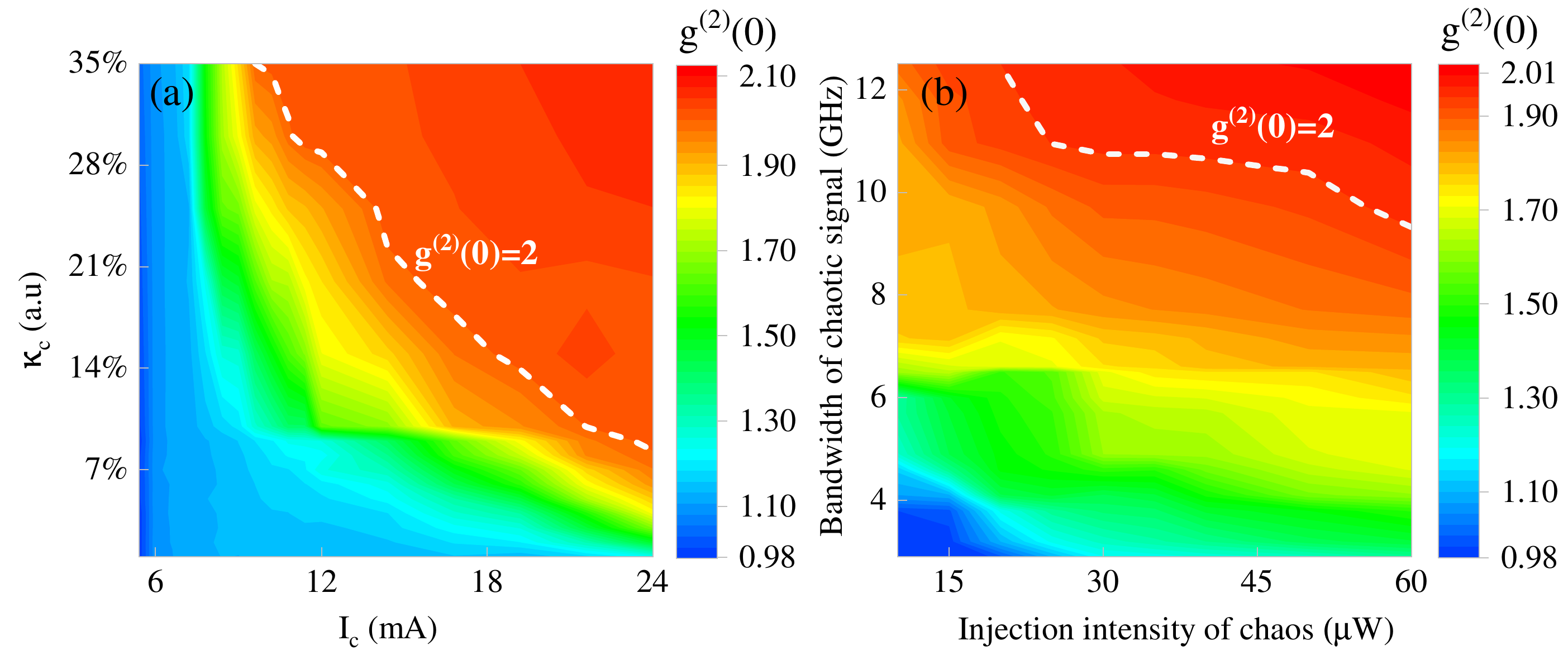}
\caption{\label{fig:3}Maps of the AQN $g^{(2)}(0)$ versus (a) feedback strengths $\kappa_{c}$ and bias current $I_{c}$, and (b) injection intensity and effective bandwidth of chaos.}
\end{figure}

To further investigate the $g^{(2)}(0)$ of the AQN, we measure high resolution maps of the ANQ $g^{(2)}(0)$ versus control and injection parameters of the chaotic laser in detail. Fig.~\ref{fig:3}(a) shows the map of the ANQ $g^{(2)}(0)$ as functions of feedback strengths  $\kappa_{c}$ and  and bias current $I_{c}$ of the chaotic lasers. In our cases of the chaos injection, the AQN $g^{(2)}(0)$ gradually reaches and even exceeds 2 as the $\kappa_{c}$ and $I_{c}$ increases. It's worth noting that the AQN behaves super-bunching effect ($g^{(2)}>2$) in the regime of high current and strong feedback. As can be seen in Fig.~\ref{fig:3}(b), the AQN $g^{(2)}(0)$ varies with effective bandwidth and injection intensity of the chaotic laser. For $I_{c}=13mA$, the AQN $g^{(2)}(0)$ increases as the measured effective bandwidth (according to the $80\%$ bandwidth definition) increases from 2.9 GHz to 12.51 GHz. The chaotic bandwidth is controlled by adjusting the feedback strength. The injection intensity of the chaotic laser with increasing from $0.5 \mu W$ to $60 \mu W$ also leads to the increasement of the AQN $g^{(2)}(0)$. The results indicate that the injection intensity and effective bandwidth of the weak chaotic signal increases is beneficial to the enhancement of the photon bunching effect of the AQN.

\begin{figure}[htbp]
\includegraphics[width=0.45\textwidth]{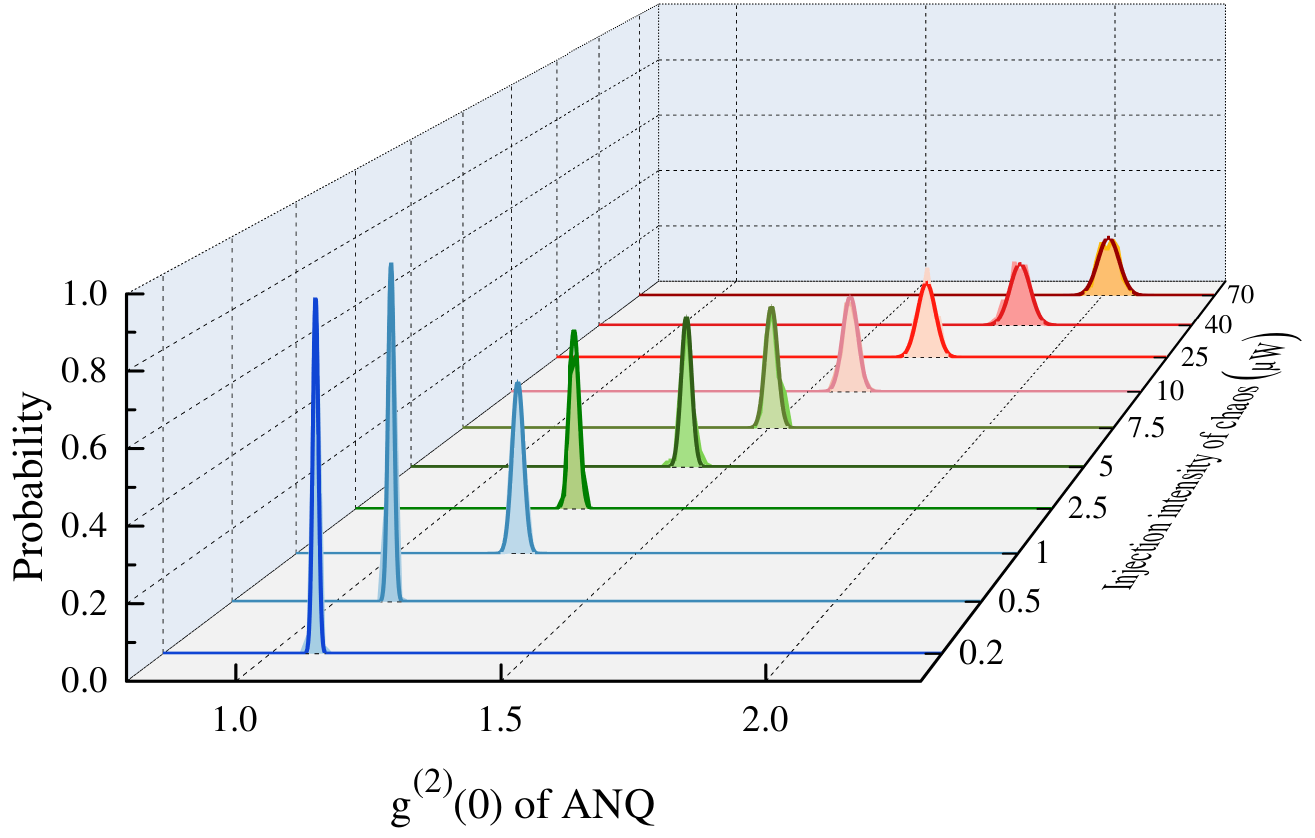}
\caption{\label{fig:4}Distributions of the AQN $g^{(2)}(0)$ for various injection intensities of chaos.}
\end{figure}

For different injection intensities of the chaos, the measured values of the AQN $g^{(2)}(0)$ exhibit probability statistical distributions. Fig.~\ref{fig:4} shows the relative frequency histograms of the AQN $g^{(2)}(0)$ for various chaotic injection intensities. When the injection bandwidth of the chaotic laser is 9.72 GHz, the $g^{(2)}(0)$ distribution of the AQN undergoes a transition from narrow Gaussian distribution to broad Gaussian distribution with increasing the injection intensity of chaos. When the injection intensity of chaos is $0.2 \mu W$, the distribution of the AQN $g^{(2)}(0)$ has a narrow and tall Gaussian shape, and is centered on $g^{(2)}(0)=1.2$. When the injection intensity of chaos is $70 \mu W$, the distribution of the AQN $g^{(2)}(0)$ has a broad Gaussian shape and is centered on $g^{(2)}(0)=2$. The distributions allow us to gain a more detailed understanding of the $g^{(2)}(0)$ of the amplified quantum noise and the relative strengths, which can be a useful tool to characterize the effect of chaos amplifying quantum noise.

In order to achieve high-speed dynamic measurement of photon correlation, we develop a photon correlation convolutional neural network (PCCNN) to speed up the processing of massive experimental data required to obtain the $g^{(2)}(0)$. The PCCNN is an improved CNN specifically designed for accelerate the photon correlation processing, and can automatically learn the features of $g^{(2)}(0)$ acquired from quadrature fluctuations. This model includes a seven-layer network structure and integrates a speed-up PyTorch library. Based on the Eq.~\ref{eq5}, the $g^{(2)}(0)$ of the AQN is derived from the measured quadrature fluctuations. We accelerate the experiment of the AQN $g^{(2)}(0)$ measurement by using PCCNN accelerating correlation data using a few quadrature fluctuations. The 30,500 sets of quadrature fluctuation signals are collected as input, among which 24,392 sets are used as training sets to train the PCCNN model. Each set contains 5,000 feature values, and the corresponding $g^{(2)}(0)$ is output. The deep-learning parameters of the PCCNN model are optimized and hundreds of sets of data are selected for the verification set. In the testing stage, we use a new dataset containing 6107 sets of data to test the generalization performance of the trained neural network, that is, how well the model performs on unseen data. A loss function is used to measure the difference between the training $g^{(2)}(0)$ of the model and the measured $g^{(2)}(0)$ of the experiment. For minimizing the error between the training and testing sets to improve the model's generalization ability, we use the Adam optimizer and a mean squared error (MSE) loss function for optimization. When the epoch reaches 200, the loss tends to converge and reaches 0.002. Finally, we accelerate the processing of 6107 sets of $g^{(2)}(0)$ in 22 seconds deploying the PCCNN. In our experiment, a million-level quadrature amplitudes are required to accurately determine $g^{(2)}(0)$. Without PCCNN deep-learning acceleration, it takes at least 4 days to calculate 6107 sets of $g^{(2)}(0)$ data with single-photon detection \cite{Guo181}, and it takes above 5 hours to obtain the same amount of data with homodyne detection. So based on the PCCNN method, we achieve at least three orders of magnitude acceleration in data acquisition time.

\begin{figure}[htbp]
\includegraphics[width=0.45\textwidth]{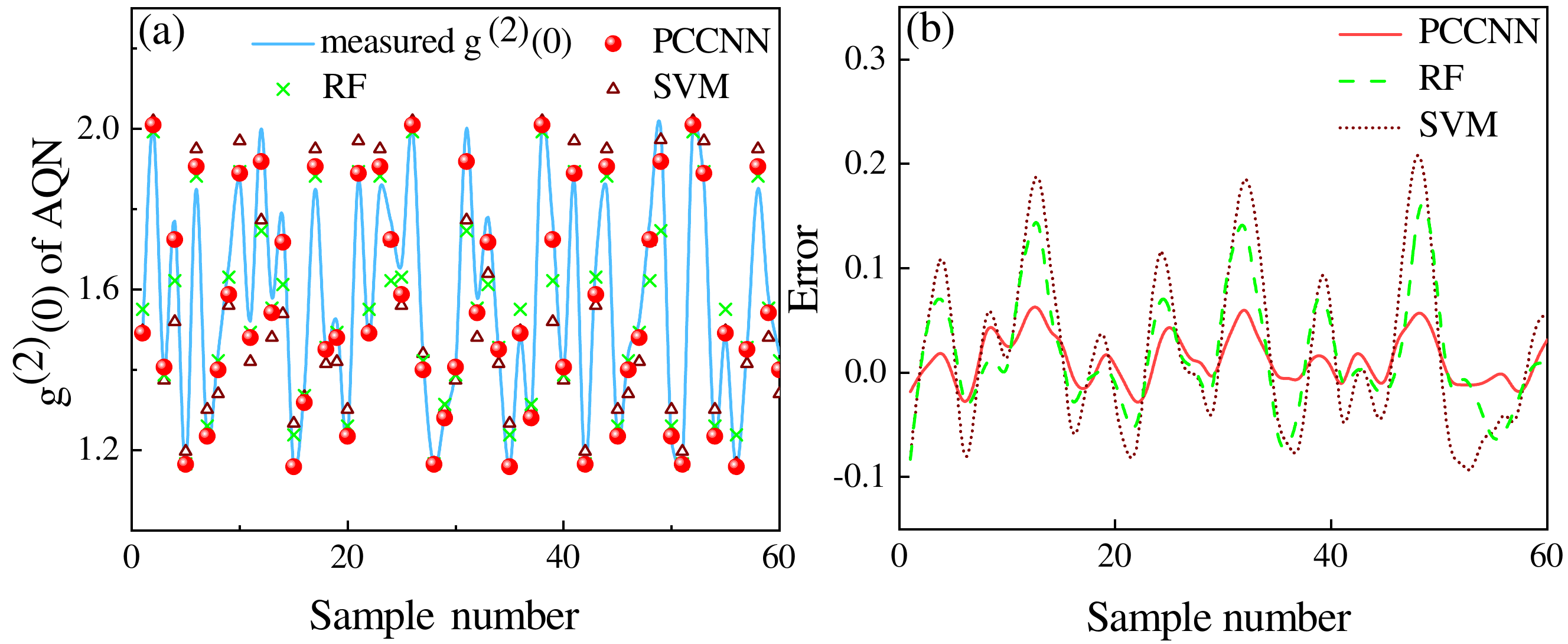}
\caption{\label{fig:5}(a) $g^{(2)}(0)$ of the AQN using PCCNN, RF and SVM accelerating methods. The blue solid line is the directly measured $g^{(2)}(0)$; (b) Errors of PCCNN, RF and SVM estimations.}
\end{figure}

We also compare the PCCNN model with two other deep-learning estimations, namely random forest (RF) and support vector machine (SVM) \cite{Navarathna21}. The results are shown in Fig.~\ref{fig:5}(a), despite using same large sets of training data, the SVM and RF models suffer from severe overfitting. The PCCNN model has the best accelerating performance due to its ability to handle large datasets and retain the original features of the data. The estimation of the PCCNN agrees well with the experiment for various experimental parameters, which indicates this deep-learning network is robust. Fig.~\ref{fig:5}(b) shows the error results of accelerating the AQN $g^{(2)}(0)$ based on the three machine-learning models. The error refers to the difference between the accelerated $g^{(2)}(0)$ and the directly measured $g^{(2)}(0)$. It can be seen that the errors in the RF and SVM acceleration are much greater than those in the PCCNN method.

\begin{figure}[htbp]
\includegraphics[width=0.45\textwidth]{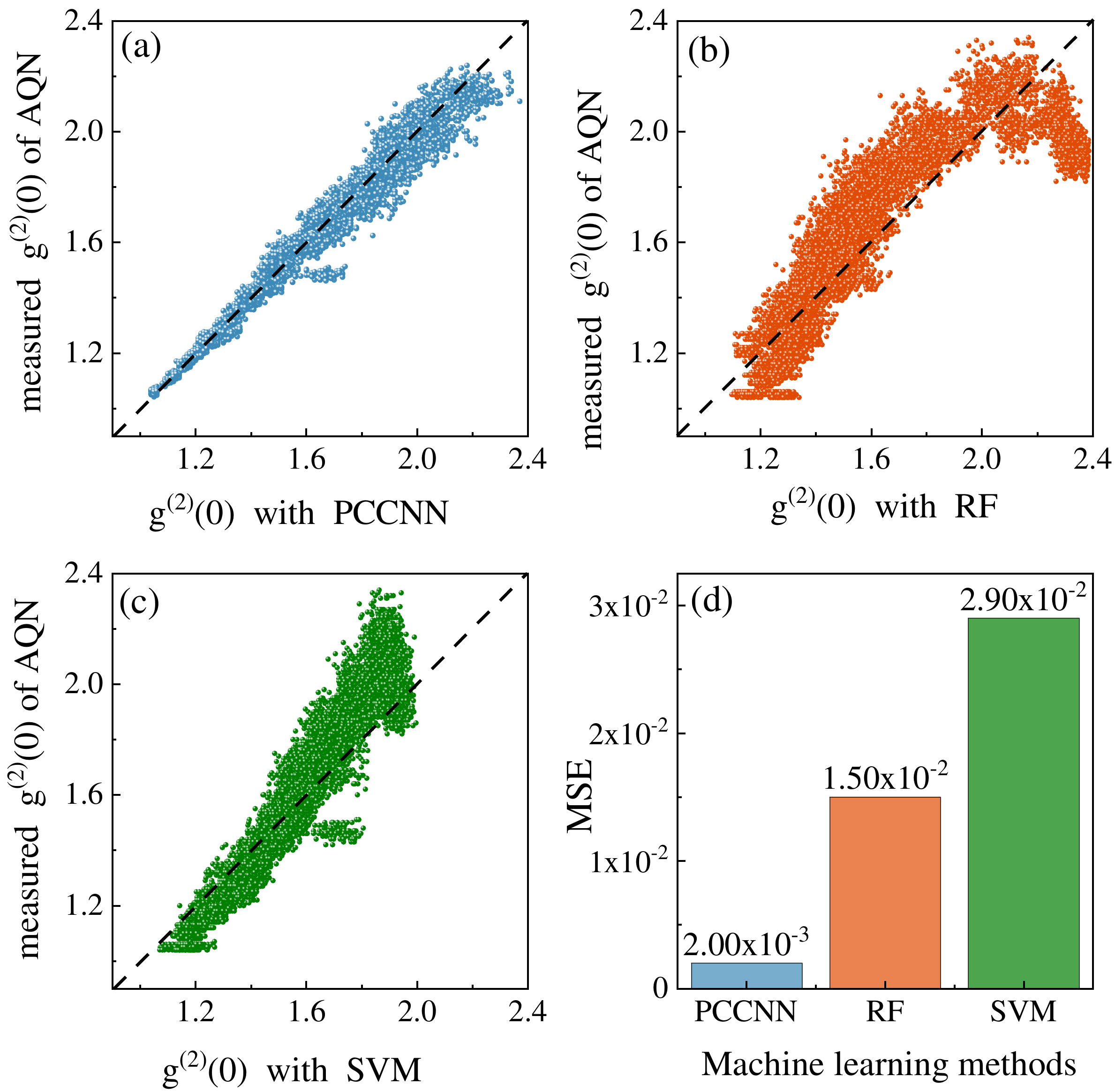}
\caption{\label{fig:6}Scatter results of the measured $g^{(2)}(0)$ vs the accelerated $g^{(2)}$(0) for the methods: (a) PCCNN, (b) RF and (c) SVM. All methods are trained using the experimental $g^{(2)}(0)$ sets. (d) MSE of PCCNN, RF and SVM.}
\end{figure}

A further analysis for the accelerating accuracy of the three learning methods is made and the scatter results are shown in Fig.~\ref{fig:6}. We use 5000 quadrature fluctuations to obtain one set of $g^{(2)}(0)$, and realize 6107 sets of $g^{(2)}(0)$ estimation by the three acceleration methods. In Figs.~\ref{fig:6}(a)-\ref{fig:6}(c), the scatter results represent the accelerated $g^{(2)}(0)$ vs the measured $g^{(2)}(0)$. The methods with high accelerating accuracy need to be aligned diagonally. The AQN $g^{(2)}(0)$ of the PCCNN method are well aligned to the diagonal line. In contrast, the RF and SVM methods tend to underestimate the $g^{(2)}(0)$ values of the AQN, and they are difficult to capture the dynamic variations of $g^{(2)}(0)$. All the three accelerating methods tackle the problem of function approximation. We evaluate the accelerating performance using mean square error (MSE), $MSE=\frac{1}{M} \sum_{m=1}^{M}\left(y_{\text {measured}}-\hat{y}_{\text {training }}\right)^{2}$. As shown in Fig.~\ref{fig:6}(d), the MSEs of PCCNN, SVM and RF are 0.002, 0.015, 0.029, respectively. The PCCNN model has a much lower MSE than the other models, and it could converge quickly with a very low loss. This indicates that the PCCNN model has high accuracy and reliability in accelerating the $g^{(2)}(0)$ of the quantum noise amplified by chaotic signal.

In conclusion, the second-order photon correlation $g^{(2)}(0)$ of the AQN are investigated experimentally based on 1.4 GHz wideband balanced homodyne detection. The AQN is prepared by injecting weak chaotic laser to amplify the quantum shot noise. For various feedback strengths, bias currents, effective bandwidths and injection intensities of the chaotic injection signal, we measure the AQN $g^{(2)}(0)$ which behaves dynamic changes from coherent state to thermal state. Furthermore, we develop a deep-learning PCCNN to process 6107 sets of the $g^{(2)}(0)$ based on quadrature fluctuations in parallel and the acceleration acquisition is achieved in 22 seconds, which is three orders of magnitude faster than the acquisition time without the deep-learning method. Compared to the other two popular machine-learning algorithms, the PCCNN method also has a high accuracy with a MSE of 0.002 in the accelerating $g^{(2)}(0)$ experiment. The technique provides an efficient and accurate solution to the problem of time-consuming photon correlation measurements and will contribute to the development and application of quantum precision measurement and imaging.

The authors thank Prof. Marc Assmann for helpful discussions during the planning stage of optical measurements. This work was supported by the National Natural Science Foundation of China (Grant Nos. 62175176 and 62075154) and the Key Research and Development Program of Shanxi Province (International Cooperation, Grant No.201903D421049).

\end{document}